\documentclass[superscriptaddress,aps,preprintnumbers,amsmath,amssymb,prd,nofootinbib,preprint]{revtex4-1}
\pdfoutput=1
\usepackage{graphicx}
\usepackage{epstopdf}
\usepackage{dcolumn}% Align table columns on decimal point
\usepackage{bm}% bold math
\usepackage{hyperref}
\usepackage{color}
\usepackage{amsmath,amssymb}
\usepackage[sort&compress]{natbib}
\usepackage{float}

\numberwithin{equation}{section}

\makeatletter
\renewcommand{\p@subsection}{}
\renewcommand{\p@subsubsection}{}
\makeatother

\begin{document}

%%%%%%%%%%%%%%%%%%%%%%%%%%%%%%%%%%%%%%%%%%%

\def\a{\alpha}
\def\b{\beta}
\def\c{\varepsilon}
\def\d{\delta}
\def\e{\epsilon}
\def\f{\phi}
\def\g{\gamma}
\def\h{\theta}
\def\k{\kappa}
\def\l{\lambda}
\def\m{\mu}
\def\n{\nu}
\def\p{\psi}
\def\q{\partial}
\def\r{\rho}
\def\s{\sigma}
\def\t{\tau}
\def\u{\upsilon}
\def\v{\varphi}
\def\w{\omega}
\def\x{\xi}
\def\y{\eta}
\def\z{\zeta}
\def\D{\Delta}
\def\G{\Gamma}
\def\H{\Theta}
\def\L{\Lambda}
\def\F{\Phi}
\def\P{\Psi}
\def\S{\Sigma}

\def\o{\over}
\def\beq{\begin{align}}
\def\eeq{\end{align}}
\newcommand{\gsim}{ \mathop{}_{\textstyle \sim}^{\textstyle >} }
\newcommand{\lsim}{ \mathop{}_{\textstyle \sim}^{\textstyle <} }
\newcommand{\vev}[1]{ \left\langle {#1} \right\rangle }
\newcommand{\bra}[1]{ \langle {#1} | }
\newcommand{\ket}[1]{ | {#1} \rangle }
\newcommand{\EV}{ {\rm eV} }
\newcommand{\KEV}{ {\rm keV} }
\newcommand{\MEV}{ {\rm MeV} }
\newcommand{\GEV}{ {\rm GeV} }
\newcommand{\TEV}{ {\rm TeV} }
\newcommand{\1}{\mbox{1}\hspace{-0.25em}\mbox{l}}
\newcommand{\headline}[1]{\noindent{\bf #1}}
\def\diag{\mathop{\rm diag}\nolimits}
\def\Spin{\mathop{\rm Spin}}
\def\SO{\mathop{\rm SO}}
\def\O{\mathop{\rm O}}
\def\SU{\mathop{\rm SU}}
\def\U{\mathop{\rm U}}
\def\Sp{\mathop{\rm Sp}}
\def\SL{\mathop{\rm SL}}
\def\tr{\mathop{\rm tr}}
\def\mpl{M_{\rm Pl}}

\def\IJMP{Int.~J.~Mod.~Phys. }
\def\MPL{Mod.~Phys.~Lett. }
\def\NP{Nucl.~Phys. }
\def\PL{Phys.~Lett. }
\def\PR{Phys.~Rev. }
\def\PRL{Phys.~Rev.~Lett. }
\def\PTP{Prog.~Theor.~Phys. }
\def\ZP{Z.~Phys. }

\def\dd{\mathrm{d}}
\def\ff{\mathrm{f}}
\def\BH{{\rm BH}}
\def\inf{{\rm inf}}
\def\ev{{\rm evap}}
\def\eq{{\rm eq}}
\def\SM{{\rm sm}}
\def\Mpl{M_{\rm Pl}}
\def\GeV{{\rm GeV}}
\newcommand{\Red}[1]{\textcolor{red}{#1}}
\newcommand{\RC}[1]{\textcolor{blue}{\bf RC: #1}}

\title{
Axion Misalignment Driven to the Hilltop 
}
\preprint{LCTP-18-33}

\author{Raymond T. Co}
\affiliation{Leinweber Center for Theoretical Physics, Department of Physics, University of Michigan, Ann Arbor, Michigan 48109, USA}
\author{Eric Gonzalez}
\affiliation{Leinweber Center for Theoretical Physics, Department of Physics, University of Michigan, Ann Arbor, Michigan 48109, USA}
\author{Keisuke Harigaya}
\affiliation{School of Natural Sciences, Institute for Advanced Study, Princeton, NJ 08540, USA}

\begin{abstract}
The QCD axion serves as a well-motivated dark matter candidate and the misalignment mechanism is known to reproduce the observed abundance with a decay constant $f_a \simeq \mathcal{O}(10^{12})$~GeV for a misalignment angle $\theta_{\rm mis} \simeq \mathcal{O}(1)$. While $f_a \ll 10^{12}$~GeV is of great experimental interest, the misalignment mechanism requires the axion to be very close to the hilltop, i.e.~$\theta_{\rm mis} \simeq \pi$. This particular choice of $\theta_{\rm mis}$ has been understood as fine-tuning the initial condition. We offer a dynamical explanation for $\theta_{\rm mis} \simeq \pi$ in a class of models. The axion dynamically relaxes to the minimum of the potential by virtue of an enhanced mass in the early universe. This minimum is subsequently converted to a hilltop because the CP phase of the theory shifts by $\pi$ when one contribution becomes subdominant to another with an opposite sign. We demonstrate explicit and viable examples in supersymmetric models where the higher dimensional Higgs coupling with the inflaton  naturally achieves both criteria. Associated phenomenology includes a strikingly sharp prediction of $3 \times 10^9~\GEV \lesssim f_a \lesssim 10^{10}$~GeV and the absence of isocurvature perturbation.
\end{abstract}

\date{\today}

\maketitle

\section{Introduction}
\label{sec:intro}
Measurements of the neutron electric dipole moment indicate an unnaturally small value of the CP-violating QCD ${\theta}$ parameter~\cite{Crewther:1979pi,Baker:2006ts}, which is known as the strong CP problem~\cite{tHooft:1976rip}. Shortly after recognizing this discrepancy, the Peccei-Quinn (PQ) mechanism~\cite{Peccei:1977hh,Peccei:1977ur} was developed as a resolution; an anomalous $U(1)_{\text{{PQ}}}$ symmetry is spontaneously broken at a scale $f_a$ and the resulting pseudo Nambu-Goldstone mode, called the axion $a$, dynamically relaxes $\bar{\theta}=\theta-\langle a/f_a\rangle$ to a vanishing value consistent with experiments.
The value of $f_a$ plays a critical role in observables related to the axion so its precise determination, theoretically and experimentally, is crucial.

A model with a weak scale decay constant was initially proposed~\cite{Weinberg:1977ma,Wilczek:1977pj} but immediately ruled out by laboratory searches. Today, supernovae cooling is the most competitive lower bound giving $f_a \gtrsim 10^8$~GeV~\cite{Ellis:1987pk,Raffelt:1987yt,Turner:1987by,Mayle:1987as,Raffelt:2006cw}. Since the axion is very light and stable on cosmological time scales, one can imagine a scenario where its relic abundance accounts for the observed dark matter (DM) abundance $\Omega_{\text{DM}}h^2=0.12$.
A thermal axion relic abundance is too hot and scarce to be consistent with cold DM. Two non-thermal production mechanisms are commonly considered.
The relic abundance from the misalignment mechanism \cite{Preskill:1982cy,Abbott:1982af,Dine:1982ah}, namely coherent oscillations due to an initial axion field value $\theta_{\rm mis}f_a$, is
\begin{equation}
\Omega_{\rm mis} h^2 \simeq 0.12~\theta_{\rm mis}^2 \mathcal{F}(\theta_{\rm mis}) \left(\frac{f_a}{5\times10^{11}~\text{GeV}}\right)^{7/6},
\end{equation}
where $\mathcal{F}(\theta_{\rm mis})$ is the anharmonicity factor. With the natural assumption of $\mathcal{O}(1)$ initial misalignment, $f_a =  10^{11}-10^{12}$ GeV is compatible with the observed DM abundance. If the PQ symmetry is broken after inflation and the domain wall number is unity, the abundance of axions emitted from the string-domain wall network is~\cite{Davis:1986xc,Kawasaki:2014sqa,Klaer:2017ond}
\begin{equation}
\label{eq:strDW}
\Omega_{\rm string,DW} h^2 \simeq 0.04-0.3 \left(\frac{f_a}{10^{11}~\text{GeV}}\right)^{7/6}.
\end{equation}
The decay constant $f_a \sim  10^{11}$ GeV reproduces the DM abundance.\footnote{The abundance is estimated assuming a scaling law, with the uncertainty given by that of the spectrum of axions. Recent studies suggest that the number of strings per horizon may increase logarithmically in time~\cite{Gorghetto:2018myk,Kawasaki:2018bzv,Martins:2018dqg}. If this is actually the case, the abundance may be larger than Eq.~(\ref{eq:strDW}).}

As ongoing axion experiments are about to reach sensitivity required to probe small decay constants of $10^8$~GeV~$<f_a<10^{12}$~GeV~\cite{Vogel:2013bta, Armengaud:2014gea, Anastassopoulos:2017kag, Rybka:2014cya, TheMADMAXWorkingGroup:2016hpc, Arvanitaki:2014dfa, Geraci:2017bmq, Sikivie:2014lha, Arvanitaki:2017nhi, Baryakhtar:2018doz,Du:2018uak}, exploring the theoretical landscape pertaining to small $f_a$ is important. Some studies have been successful in allowing small $f_a$ in a natural setting, such as parametric resonance from a PQ symmetry breaking field~\cite{Co:2017mop} and decays of quasi-stable domain walls~\cite{Hiramatsu:2010yn,Hiramatsu:2012sc,Kawasaki:2014sqa,Harigaya:2018ooc}.
The misalignment mechanism can reproduce the observed DM abundance for $f_a \ll  10^{12}$~GeV if $\theta_{\rm mis}$ is taken sufficiently close to $\pi$ \cite{Turner:1985si, Lyth:1991ub, Strobl:1994wk, Bae:2008ue, Visinelli:2009zm}, where the anharmonicity factor $\mathcal{F}(\theta_{\rm mis})$ becomes important.

In this study, we propose a scenario which dynamically predicts $\theta_{\rm mis} \simeq \pi$ and thus small $f_a$ in the context of axion DM from the misalignment mechanism. It is commonly assumed that no misalignment angles are special in the early universe, and $\theta_{\rm mis} \simeq \pi$ requires a fine-tuned initial condition.
This is not the case given two conditions are met: 1) the axion field dynamically relaxes to the minimum of the potential in the early universe and 2) the model possesses a non-trivial prediction between the minima of the axion potential in the early and today's epochs.
We refer to the axion relaxation with the fulfillment of these requirements as Dynamical Axion Misalignment Production (DAMP). We study DAMP by the dynamics of the Higgs fields during inflation. The mechanism follows from suspending the assumption that axion's late-time dynamics is agnostic to inflationary dynamics. To be concrete, we study the Minimal Supersymmetric Standard Model (MSSM). The Higgs fields $H_u$ and $H_d$ in general couple to the inflaton potential energy via higher dimensional operators, which lead to so-called Hubble induced masses.
The Higgs fields can acquire a large field value in the early universe by virtue of the Hubble induced mass.
This large field value gives large quark masses, which enhance the confinement scale to $\Lambda_{\text{QCD}}'$ during inflation. Since $m_a$ is proportional to $\Lambda_{\text{QCD}}'$, this raises the axion mass to allow for earlier relaxation to the minimum. Note that we need to assume the Higgs fields are not charged under PQ symmetry; otherwise, the decay constant will be as high as the Higgs VEV and suppress the axion mass.
For context, early studies~\cite{Dvali:1995ce,Banks:1996ea,Choi:1996fs} have made use of moduli fields to raise the QCD confinement scale $\Lambda_{QCD}\rightarrow\Lambda_{QCD}'$ during inflation. 
This avoids fine-tuning problems that arise under the assumption of an $\mathcal{O}(1)$ initial misalignment with large values $f_a>10^{12}$ GeV. Later studies \cite{Jeong:2013xta,Choi:2015zra} used Higgs fields as the moduli fields and refined the scope of the mechanism to reduce isocurvature perturbations for models with large inflation scales, which comes at the cost of an inability to suppress the axion abundance. This loss of abundance predictability is because no assumptions are made about the evolution of the axion minimum through inflation. In the MSSM for example, we have~\cite{Choi:1996fs}
\begin{equation}
\label{eq:theta_eff}
\theta_{\text{eff}}=\theta_\text{QCD} + \arg(\det\lambda_u\lambda_d)+3 \arg(m_{\tilde{g}})+3\arg(B\mu) ,
\end{equation}
where $\lambda_u, \lambda_d$ are the Yukawa coupling matrices, $m_{\tilde{g}}$ denotes the gluino mass, and $B\mu$ is the soft breaking mass for Higgs scalars.
Although a large $\Lambda_\text{QCD}'$ can help fulfill the first DAMP criterion, we should also explain how a Hubble induced mass fits in with Eq~(\ref{eq:theta_eff}) to fulfill the second criterion.

The K\"ahler potential can give rise to a Hubble induced $B\mu$ term. If the argument of the term is different from the vacuum $B\mu$ term by $\pi$ and dominates, a shift of $\pi$ relative to the vacuum value is induced in the axion potential.\footnote{If the arguments are the same, we may dynamically relax the axion to today's minimum during inflation as discussed in a separate paper \cite{Co:2018phi}.} The difference of $\pi$ in the arguments can be understood by the (approximate) CP symmetry of the theory, such that the $B\mu$ terms are real and the difference of $\pi$ is simply the opposite signs of the terms. An approximate CP symmetry is also invoked in Ref.~\cite{Banks:1996ea}, where a relaxation to $\theta_{\rm mis} \simeq 0$ is considered. Note that the shift of $\pi$ in the axion potential occurs only if the number of generations is odd.

A large $m_a$ allows the axion field to relax to the bottom of the potential during inflation, and a $\pi$ shifted axion potential means this minimum coincides with today's hilltop. Without additional particles beyond the MSSM, $\Lambda_\text{QCD}'$ and consequently $m_a$ cannot be arbitrarily large; we find $m_a \lesssim 10 $ TeV. Thus, in the minimal scenario, we consider TeV scales for Hubble during inflation $H_I$ to allow for the relaxation of the axion misalignment during inflation. We also explore non-minimal models where $m_a$ and thus $H_I$ can be larger. Relaxing the axion arbitrarily close to today's hilltop may cause overproduction of axion DM, but we find that the running of Yukawa terms in the Standard Model (SM) gives a sufficient CP phase change $\mathcal{O}(10^{-16})$ to avoid the scenario \cite{Ellis:1978hq, Khriplovich:1981ca}. An exciting implication of this mechanism is that $f_a$ is fixed to roughly $3\times10^9$~GeV by the observed DM abundance and CP-violating phase renormalization in the theory. We impose CP symmetry in the Higgs and inflaton sectors. Additional CP violation (CPV) of up to $\mathcal{O}(10^{-4})$ only induces $\mathcal{O}(1)$ changes in the prediction of $f_a$. In summary, by the inflationary dynamics of the Higgs fields as well as the (approximate) CP symmetry, we can fulfill both criteria of a DAMP scenario; in particular in this paper we explore the case where the inflationary minimum is shifted by $\pi$ from today's minimum, which is referred to as DAMP$_\pi$.

We now elaborate on the approximate CP symmetry. Although the $\mathcal{O}(1)$ amount of CPV measured in the SM must be generated in the theory, a small CPV in the extended sectors can be a consequence of the suppressed couplings with the source of CP violation. Such hierarchical couplings can result from the protection of additional symmetries or the geometric separation in the extra dimensions. Additionally, any quantum corrections that attempt to transfer $\mathcal{O}(1)$ CPV from the SM to the extended sectors are automatically small. The reason is that the CP phase of the Yukawa couplings only becomes physical when all three generations are involved, suggesting that the interactions are suppressed by small Yukawa couplings, mixing among generations, and higher loop factors.%
\footnote{Even though the CP symmetry is a solution to the strong CP problem alternative to the axion, the $\mathcal{O}(1)$ CPV in the Yukawa sector may unacceptably modify the $\theta$ term. For models that avoid such consequences, refer to Refs.~\cite{Nelson:1983zb,Barr:1984qx,Bento:1991ez,Hiller:2001qg}}
With this CP structure, the implications for the extended sectors are as follows. The (approximate) CP symmetry treats the CP-odd and CP-even moduli differently in such a way that the moduli affecting the axion minimum can be stabilized at the CP-conserving points. The smallness of the CP phases is also guaranteed in the masses of any additional colored particles that we introduce in the non-minimal models. Crucially, a CP symmetry only ensures all relevant parameters are real but does not forbid the change of signs throughout the evolution; this is exactly what can give rise to a shift of $\pi$ in the axion potential.

In Sec.~\ref{sec:misalign} we briefly review the axion misalignment mechanism and the role of the anharmonicity factor. We also discuss how the amount of CPV in a theory can be connected to the axion abundance in a DAMP$_\pi$ model. In Sec.~\ref{sec:model} we show how a Hubble induced mass for the Higgs in the early universe can induce an axion mass enhancement and a phase shift of $\pi$ in the axion potential, fulfilling the DAMP$_\pi$ criteria. In Sec.~\ref{sec:cosmo} we discuss both a set of minimal models with the cosmology fully evaluated, and extended models with a larger viable parameter space and a simplified discussion of the post-inflationary cosmology. Finally, in Sec.~\ref{sec:conclusion} we summarize and discuss the possible phenomenological implications of this model as well as future directions.

\section{Axion Misalignment \& Early Relaxation}
\label{sec:misalign}
We first review the axion misalignment mechanism. The equation of motion and energy density of axions are given by 
\begin{align}
\label{eq:EoM_a}
\ddot{\theta}_a + 3 H \dot{\theta}_a & = -m_a^2 \sin \theta_a \\
\label{eq:rho_a}
\rho_a & = \frac{1}{2} \left(m_a^2 \, a^2 + \dot{a}^2 \right) 
\end{align}
where $\theta_a \equiv a / f_a$ parametrizes the axion field value $a$ and $H$ is the Hubble expansion rate. In principle, all relevant parameters in Eqs.~(\ref{eq:EoM_a})~and~(\ref{eq:rho_a}) can vary throughout cosmological evolutions but we focus on the case where the PQ symmetry is already broken during inflation. The misalignment contribution to the axion DM abundance is as follows. In the conventional setup where $m_a$ is assumed negligible compared to the Hubble parameter during inflation, the axion field value is practically frozen due to a large Hubble friction term in Eq.~(\ref{eq:EoM_a}) with the solution approximated by
\begin{equation}
\label{eq:theta_fric}
\theta_0 \simeq \theta_i e^{- N_e\,m_a^2/3 H_I^2 }  \quad \quad \text{for} \quad m_a \ll H_I ,
\end{equation}
with $\theta_0$ ($\theta_i$) the angle at the end (onset) of inflation, unless the number of e-folding is exceedingly large $N_e \sim (H_I/m_a)^2$ as pointed out by Refs.~\cite{Dimopoulos:1988pw,Graham:2018jyp, Guth:2018hsa}. As a result of inflation, the misalignment angle takes a random but uniform value $\theta_{\rm mis}$ in the observable universe. Around the QCD phase transition, the axion acquires a mass from the QCD non-perturbative effects and starts to oscillate, when $3 H \simeq m_a$, from $a_{\rm mis} = \theta_{\rm mis}f_a$ towards the minimum today. Without fine-tuning, $\theta_{\rm mis}$ is expected to be order unity.  The coherent oscillations of axions contribute to the cold dark matter abundance 
\begin{equation}
\label{eq:omega_a}
\Omega_a h^2 = 0.12 \langle \theta_{\rm mis}^2 \mathcal{F}(\theta_{\rm mis}) \rangle \left(\frac{f_a}{5\times10^{11}\text{GeV}}\right)^{7/6} ,
\end{equation}
where $\mathcal{F}(\theta_{\rm mis}) \simeq 1$ for $\theta_{\rm mis} \ll \pi$ and, for $\theta_{\rm mis} \gtrsim 0.9 \, \pi$, is analytically approximated by \cite{Lyth:1991ub}
\begin{equation}
\label{eq:F_anh}
\mathcal{F}(\theta_{\rm mis}) \simeq \frac{16 \sqrt{2}}{\pi^3} \left[ \ln \left(\frac{1}{1-\theta_{\rm mis}/\pi}\right)\right]^{7/6}.
\end{equation}
Several numerical studies have been devoted to the determination of $\mathcal{F}(\theta_{\rm mis})$ \cite{Turner:1985si,Strobl:1994wk,Bae:2008ue,Wantz:2009it} but DAMP$_{\pi}$ calls for a dedicated study for the extreme limit of $\pi - \theta_{\rm mis} \ll 1$. 
The exponents in Eqs.~(\ref{eq:omega_a})~and~(\ref{eq:F_anh}) assume the topological susceptibility of QCD given by the dilute instanton gas approximation (see the lattice results in Refs.~\cite{Petreczky:2016vrs, Borsanyi:2016ksw, Burger:2018fvb, Bonati:2018blm, Gorghetto:2018ocs}) but our results are insensitive to this uncertainty.

\begin{figure}[t]
	\includegraphics[width=\linewidth]{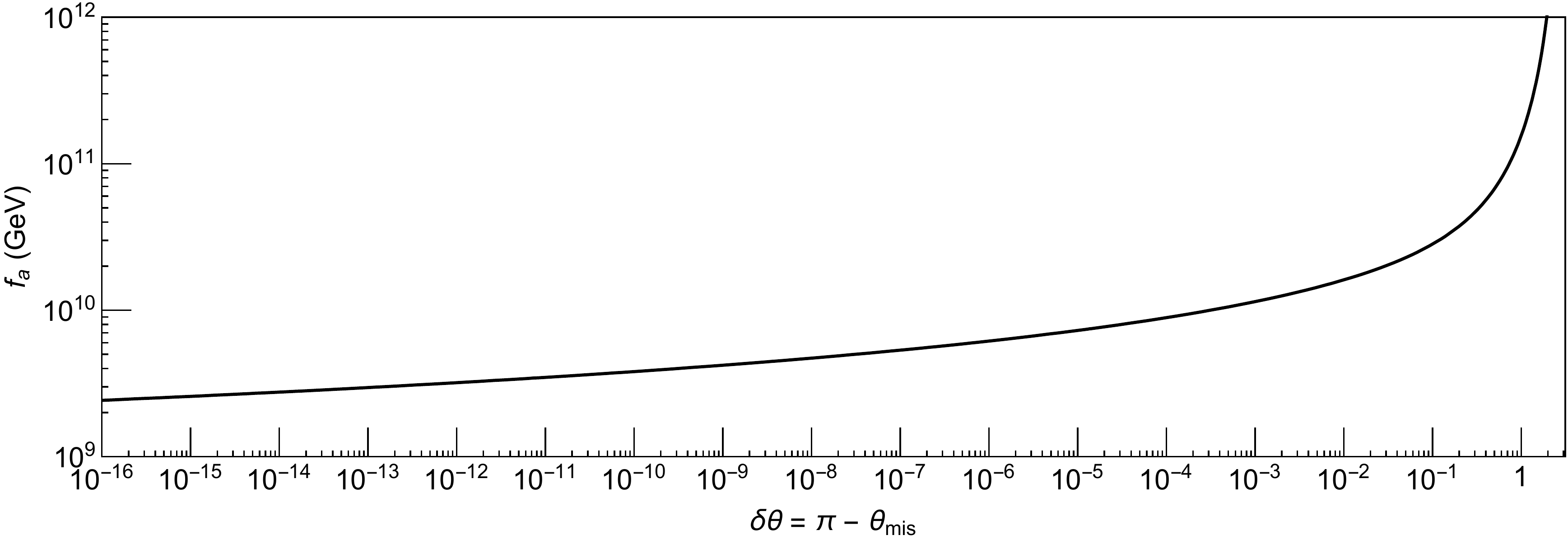}
	\caption{Insensitivity of $f_a$ to very small shifts from the hilltop.}
	\label{fig:fa}	
\end{figure}
We now discuss how our framework, by relaxing the above assumptions, makes a prediction for $f_a$ using the DM abundance and the CP-violating phase $\delta\theta_{\rm CP}$ of the theory. The axion mass arises from QCD dynamics. There is no a priori reason that the axion mass during inflation is given by the exact same QCD effect observed today. In fact, there are numerous scenarios where the axion is enhanced in the early universe, e.g.~a large QCD confinement scale \cite{Dvali:1995ce,Banks:1996ea,Choi:1996fs, Jeong:2013xta,Choi:2015zra,Ipek:2018lhm}, explicit PQ breaking \cite{Higaki:2014ooa, Kawasaki:2015lea, Takahashi:2015waa, Kearney:2016vqw}, and magnetic monopoles \cite{Kawasaki:2015lpf,Nomura:2015xil}. If the axion mass is larger than the Hubble scale during inflation, the axion starts oscillations and is rapidly relaxed towards the minimum,
\begin{equation}
\label{eq:theta_relax}
\theta_0 \simeq \theta_i e^{-3 N_e/2}  \quad \quad \text{for} \quad m_a \gg H_I .
\end{equation}
Additionally, if the CP phase of the model has a phase shift of $\pi$ after inflation, as explained in Sec.~\ref{sec:intro} and elaborated in Sec.~\ref{sec:model}, the location of this minimum is then converted into the maximum of the potential, making the effective misalignment angle $\theta_{\rm mis} \simeq \pi$. There is however a limit on how close $\theta_{\rm mis}$ can be to $\pi$ because the quantum correction to the $\theta$ parameter from the CP violation in the SM Yukawa couplings is $\delta\theta_{\rm CP} \sim 10^{-16}$ \cite{Ellis:1978hq, Khriplovich:1981ca} and the running of the Yukawa couplings necessarily induces a phase shift of similar order between the inflationary and low energy scales. 
This small deviation from the hilltop $\delta\theta = \pi - \theta_{\rm mis}$ allows for the prediction of $f_a$ due to the anharmonic effects. In the limit $\theta_{\rm mis} \rightarrow \pi$, $\mathcal{F}(\theta_{\rm mis})$ and thus $\Omega_a h^2$ are only logarithmic dependent on $\delta\theta$
so one can predict $f_a$ in terms of the deviation $\delta \theta$ by requiring DM abundance using Eqs.~(\ref{eq:omega_a})~and~(\ref{eq:F_anh})
\begin{equation}
\label{eq:fa_F_anh}
f_a \simeq 2.4 \times 10^{9} \ \text{GeV} \left(\frac{\Omega h^2}{0.11}\right)^{6/7}	\left(1 + 0.026 \ln \left(\frac{\delta\theta}{10^{-16}}\right) \right)  ,
\end{equation}
where we assume $ \left| \ln \left(\frac{\delta\theta}{10^{-16}}\right) \right| \ll  \ln\left(\frac{\pi}{\delta\theta_{\rm CP}}\right)  \simeq 38$ or equivalently $\delta\theta \ll 1$. This is the striking feature of the anharmonic effect---the value of $f_a$ necessary for the DM abundance has an exceptionally mild logarithmic dependence on the CP phase shift as long as it is much less than unity. The sharp prediction of $f_a$ is illustrated in Fig.~\ref{fig:fa} using the analytic approximations in Eqs.~(\ref{eq:omega_a})~and~(\ref{eq:F_anh}).
The prediction on the decay constant only changes by $\mathcal{O}(1)$ factor even if CPV of $\mathcal{O}(10^{-4})$ is added.

\section{Dynamical Axion Misalignment Production at the Hilltop}
\label{sec:model}
We would like to show that allowing $H_u$ and $H_d$ to acquire large VEVs during inflation can lead to a DAMP$_\pi$ scenario. To guide the reader, we first restate the conditions under which the DAMP model is applicable: 1) the axion field dynamically relaxes to the minimum of the potential in the early universe and 2) the model possesses a non-trivial prediction between the minima of the axion potential in the early and today's epochs. Throughout our discussion of DAMP$_\pi$ models, we have in mind a minimal model as a proof of principle and extended models to further explore viable parameter space. Generically, we can include inflaton-Higgs dynamics with the effective operators suppressed by the cutoff scale $M$ in the K\"ahler potential
\begin{align}
\label{eq:Kahler}
\Delta K =  \frac{|X|^2}{M^2} \left( |H_u|^2 + |H_d|^2 - \big( H_u H_d + c.c. \big)  - \frac{|H_u|^2 |H_d|^2}{M^2}  - \frac{|H_u|^4}{M^2}  - \frac{|H_d|^4}{M^2}  \right),
\end{align}
where $X$ is the chiral field whose $F$-term provides the inflaton potential energy. We omit $\mathcal{O}(1)$ coupling constants here and hereafter.
For illustration purposes, we only show lower dimensional operators relevant for the following discussion. Higher dimensional operators do not change the discussion.

During inflation, the inflaton $F$-term gives the Higgs fields Hubble induced terms,
\begin{align}
\label{eq:HiggsL}
\Delta V =  c H_I^2 \left( - |H_u|^2 - |H_d|^2   + \big( H_u H_d + c.c. \big) 
 + \frac{|H_u|^2|H_d|^2}{M^2}  +  \frac{ |H_u|^4 }{M^2}  +  \frac{ |H_d|^4 }{M^2} \right), 
\end{align}
where $c=(\mpl/M)^2$ and $H_I$ is the Hubble scale during inflation.
We assume that the Hubble induced mass terms are negative. They push the Higgs fields in the $D$-flat direction $|H_u| = |H_d|$ up to the cutoff scale $M$, and as we will see in the following sections the large Higgs VEVs realize DAMP$_\pi$.

\subsection{Axion Mass During Inflation}
Together with the effective terms from the K\"ahler potential in Eq.~(\ref{eq:Kahler}), the MSSM Higgs potential reads
\begin{equation}
\label{eq:Vhiggs}
\begin{split}
V_{\text{Higgs}} = & \left(|\mu|^2+m_{H_u}^2 - cH_I^2 \right) |H_u|^2 + \left(|\mu|^2+m_{H_d}^2 - c H_I^2 \right) |H_d|^2 - \left(B\mu - cH_I^2\right) (H_u H_d + c.c.) \\
& + \frac{g^2+g'^{2}}{8}\left(|H_u|^2-|H_d|^2\right)^2 + \frac{g^2}{2} \left| H_u H_d^* \right|^2 + \frac{cH_I^2}{M^2}\left(|H_u|^2|H_d|^2 +|H_u|^4 +|H_d|^4 \right) .
\end{split}
\end{equation}
We assume that the Higgs sector is nearly CP symmetric, which is anyway required from the limits on the electric dipole moment for TeV scale supersymmetry. See Refs.~\cite{Andreev:2018ayy} and~\cite{Cesarotti:2018huy} for the latest measurement and its implication to supersymmetric theories, respectively. We also assume a CP symmetry in the inflaton-Higgs coupling.
The Higgs fields break $SU(2)_L\times U(1)_Y\rightarrow U(1)_{\text{EM}}$ by both the early universe VEV and today's VEV.
Parameterizing the Higgs field space in terms of a radial mode $\phi \equiv |H_u| = |H_d|$ along the $D$-flat direction and an angular mode $\xi=\arg{(H_u H_d)}$, which is the relative phase of the Higgs fields, allows us to write
\begin{equation}
\label{eq:VDflat}
V \simeq (m_\text{SUSY}^2 - c H_I^2)\phi^2 - (B\mu-cH_I^2)\cos({\xi}) \phi^2+ \frac{cH_I^2}{M^2}\phi^4,
\end{equation}
where we have taken $m_{H_u}\sim m_{H_d} \sim \mu \sim m_\text{SUSY}$.
The phases of $H_uH_d$ are chosen so that $\xi =0$ in the vacuum today.
The radial mode, for $\sqrt{c}H_I\gtrsim m_\text{SUSY}$, acquires a large VEV of order $\phi_i \sim M$. This is clearly seen from minimizing the potential.
Assuming that the sign of the Hubble induced $B\mu$ term is opposite to the vacuum one as shown in Eq.~(\ref{eq:VDflat}), the phase initially obtains a value during inflation of $\xi = \pi$, while today's value is $\xi = 0$.
We discuss the implication of the phase shift in the next subsection and focus this subsection on the large radial direction.%
\footnote{In the extended model discussed below, the sign flip of the $B\mu$ is not necessary. The Hubble induced $B\mu$ term is not necessary as long as the vacuum one is larger than $H_I^2$.}

The large VEV $\phi_i$ gives quarks very large masses during inflation. In the MSSM, the 1-loop renormalization group equation (RGE) is 
\begin{equation}
	\mu_r \frac{d}{d\mu_r} \frac{8 \pi^2}{g^2} = 3N-F,
\end{equation}
where $\mu_r$ is the renormalization scale, $N=3$ is the gauge group index, and $F$ is the number of active fermions in the theory.
Solving the RGE from the TeV scale up to the scale $\phi_i$, and from the scale down while pretending that all quarks are above the scale where the gauge coupling diverges, we obtain the fiducial dynamical scale
\begin{equation}
\label{eq:LQCDhiggs}
\Lambda_\text{fid} = 10^7~{\rm GeV} \left( \frac{\phi_i}{10^{16}~{\rm GeV}} \right)^{2/3} \left( \frac{{\rm tan}\beta}{1} \right)^{1/3} .
\end{equation}
This is the physical dynamic scale $\Lambda_{\rm QCD}'$ if all quarks (including the KSVZ quarks \cite{Kim:1979if,Shifman:1979if}) are above the scale. If some quarks are below the scale, the physical dynamical scale $\Lambda_{\rm QCD}'$ is given by
\begin{equation}
\label{eq:Lfid}
\Lambda_\text{fid} = \Lambda_\text{QCD}' \times \prod_{m_q < \Lambda_\text{QCD}'} \left( \frac{m_q}{\Lambda_\text{QCD}'
} \right)^{1/9}.
\end{equation}
The axion mass vanishes when the gluino is massless since strong dynamics gives the mass dominantly to the R-axion. The axion mass is hence given by
\begin{align}
\label{eq:ma}
m_a \simeq \frac{1}{4\pi} \frac{m_{\tilde{g}}^{1/2} \Lambda_\text{fid}^{3/2}}{f_a},
\end{align}
where we assume that the gluino mass is below the physical dynamical scale and that the large Higgs VEV does not break the PQ symmetry. We include the factor of $4\pi$ expected from the naive dimensional analysis~\cite{Manohar:1983md,Georgi:1986kr,Luty:1997fk,Cohen:1997rt}.
Here $m_{\tilde{g}}$ is the RGE invariant one, $m_{\tilde{g},{\rm phys}}/g^2$.
The holomorphy of the gauge coupling guarantees that we may use the fiducial dynamical scale to evaluate the axion mass. Physically, the suppression of the fiducial dynamical scale in comparison with the physical dynamical scale takes into account the suppression of the axion mass by light quarks.
For the minimal setup where the dynamical scale is raised solely by large Higgs VEVs as in Eq.~(\ref{eq:LQCDhiggs}),
\begin{equation}
m_a \simeq 30 \, \GEV \left( \frac{m_{\tilde{g}}}{\TEV} \right)^{1/2} \left( \frac{\Lambda_\text{fid}}{10^7 \ \GEV} \right)^{3/2} \left( \frac{3 \times 10^9 \ \GEV}{f_a} \right) .
\end{equation}

We may raise the dynamical scale further by introducing additional particles. One possibility is to introduce a moduli field whose field value controls the gauge coupling \cite{Dvali:1995ce,Banks:1996ea,Choi:1996fs, Jeong:2013xta, Choi:2015zra, Ipek:2018lhm}, and assume that the moduli field value during inflation raises the gauge coupling constant. Another possibility is to introduce additional $SU(3)_c$ charged fields and assume that their masses are large during inflation as considered in Ref.~\cite{Jeong:2013xta}. A field whose field value controls the masses of the additional particles can be regarded as a moduli field.
For $N_\Psi$ pairs of $SU(3)_c$ fundamental chiral fields with a mass $M_{\Psi}$ and $M_{\Psi,I}$ in the vacuum and during inflation respectively, the dynamical scale is given by
\begin{equation}
\label{eq:LQCD}
\Lambda_\text{fid} = 10^7 \ \text{GeV}\ \left(\frac{M_{\Psi,I}}{M_\Psi}\right)^{N_\Psi/9} \left( \frac{\phi_i}{10^{16}~{\rm GeV}} \right)^{2/3} \left( \frac{{\rm tan}\beta}{1} \right)^{1/3}.
\end{equation}
To achieve the second requirement of the DAMP scenario, CPV phases in $M_{\Psi,I}$ and $M_\Psi$ should be absent. Instead of flipping the sign of the $B\mu$ tern, we may flip the sign of the masses of $\Psi$ to achieve DAMP$_\pi$.
We will see later in Sec.~\ref{sec:cosmo} that this dynamical scale cannot be arbitrarily large because of the backreaction of strong dynamics to the Higgs as well as the PQ sector.

Combining Eqs.~(\ref{eq:ma})~and~(\ref{eq:LQCD}) we find that for appropriate values of $H_I$, the early universe axion mass is large enough for relaxation of the axion field to its minimum. Since the largeness of the dynamical scale $\Lambda_\text{QCD}'$ depends on the VEV of $\phi$, the decay of the inflaton and proceeding relaxation of $\phi$ to today's VEV means that the post-inflationary cosmology is non-trivial. Prior to exploring this complex cosmology, however, we turn our attention to the relative $\pi$ phase shift of the axion potential.

\subsection{Shifted Axion Potential}
Another consequence of a large Higgs VEV during inflation from the K\"ahler potential in Eq.~(\ref{eq:Kahler}) is that the relative phase between the Higgs fields is shifted by $\pi$
as can be seen explicitly in Eq.~($\ref{eq:VDflat}$). The shift of $\xi$ also shifts the minimum of the axion potential: Eq.~(\ref{eq:theta_eff}) shows a direct connection between the $B\mu$ term and the axion misalignment angle minimum $\theta_\text{eff}$.
Once the inflaton decays or its energy is redshifted and the Hubble induced terms become subdominant, minimization of the potential is achieved for $\xi = 0$.
In the extended model discussed in the previous section, the sign flip of the masses of extra quarks $\Psi$ can achieve a similar situation.

The phase shift of the axion potential is not exactly $\pi$ because of the $\mathcal{O}(1)$ renormalization of Yukawa couplings. Since the CPV from Yukawa couplings manifests as $\mathcal{O}(10^{-16})$ shifts in the axion potential \cite{Ellis:1978hq, Khriplovich:1981ca}, running these couplings from the early large Higgs VEVs to the electroweak scale necessarily induces an $\mathcal{O}(10^{-16})$ shift in the axion potential. We may also add small CPV to the $B\mu$ terms to induce further shift.  Even if the shift is as large as $\mathcal{O}(10^{-4})$, the prediction of $f_a$ changes only by an $\mathcal{O}(1)$ factor. 

To summarize, a K\"ahler potential such as the one in Eq.~($\ref{eq:Kahler})$ gives Higgs fields Hubble induced masses, and the $D$-flat potential in Eq.~($\ref{eq:VDflat})$ is minimized at a large Higgs VEV with an opposite phase from today. The $\pi$ shifted Higgs phase $\xi$ along with the Yukawa coupling renormalization induce a shift in the axion potential by $\pi - \mathcal{O}(10^{-16})$. This sets the scene for a DAMP$_\pi$ scenario where the DM abundance is given by Eq.~(\ref{eq:omega_a}) and the value of $f_a$ can be predicted using the anharmonicity factor Eq.~(\ref{eq:F_anh}).

\section{Cosmological Evolution}
\label{sec:cosmo}

In Sec.~\ref{sec:model} we demonstrated that, in the early universe, both a large axion mass and a phase shift of $\pi$ are possible due to a large Higgs VEV with an opposite phase from today. The remaining question is whether there exists a viable cosmology with a consistent evolution between the two periods without spoiling predictions. We first explore the inflationary and post-inflationary constraints in the minimal model, and then later comment on the broader parameter space allowed by extended models.

\subsection{Minimal Models}
\label{subsec:cosmo_min}

\begin{figure}[t]
	\includegraphics[width=0.495\linewidth]{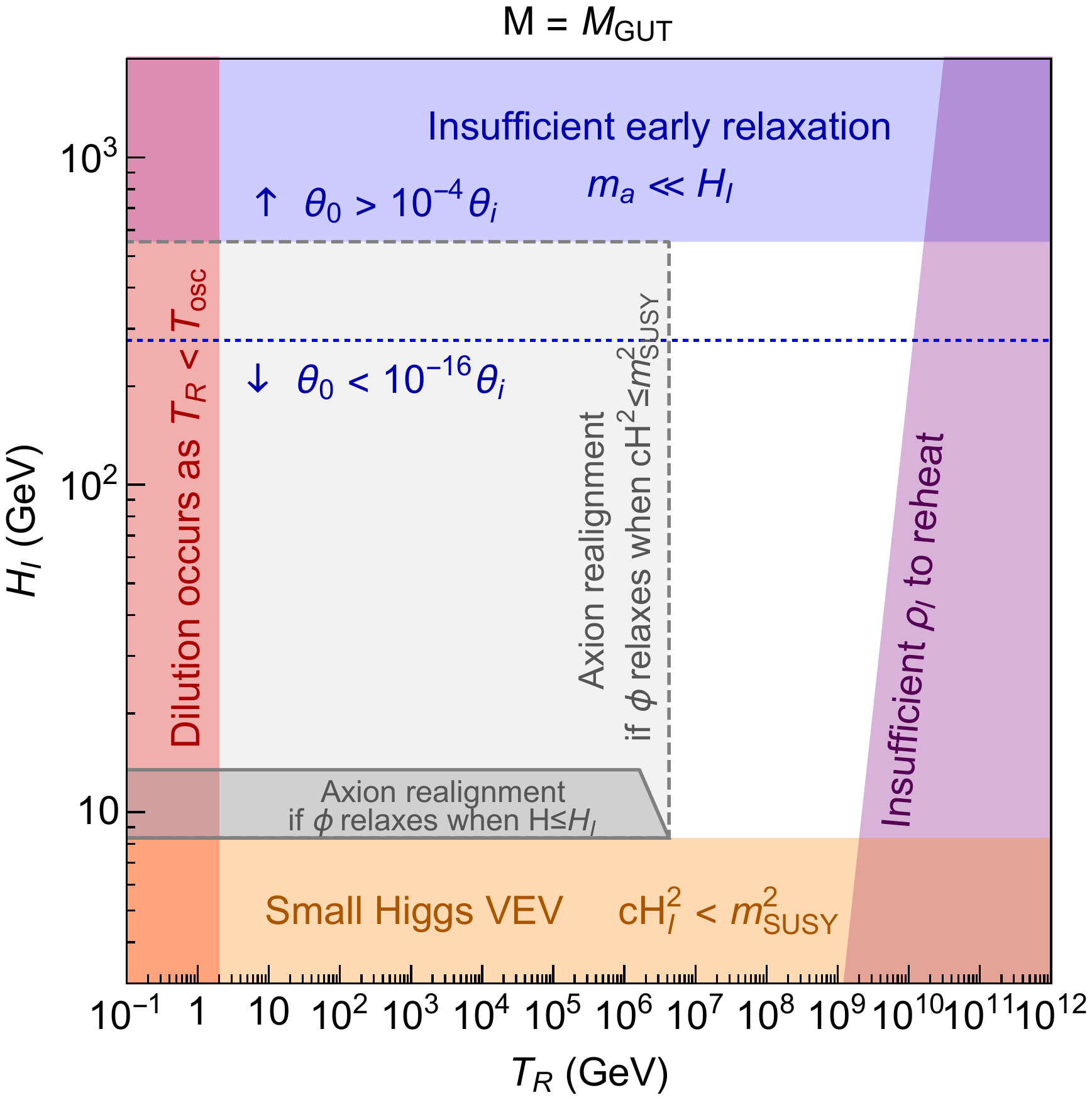}  	\includegraphics[width=0.495\linewidth]{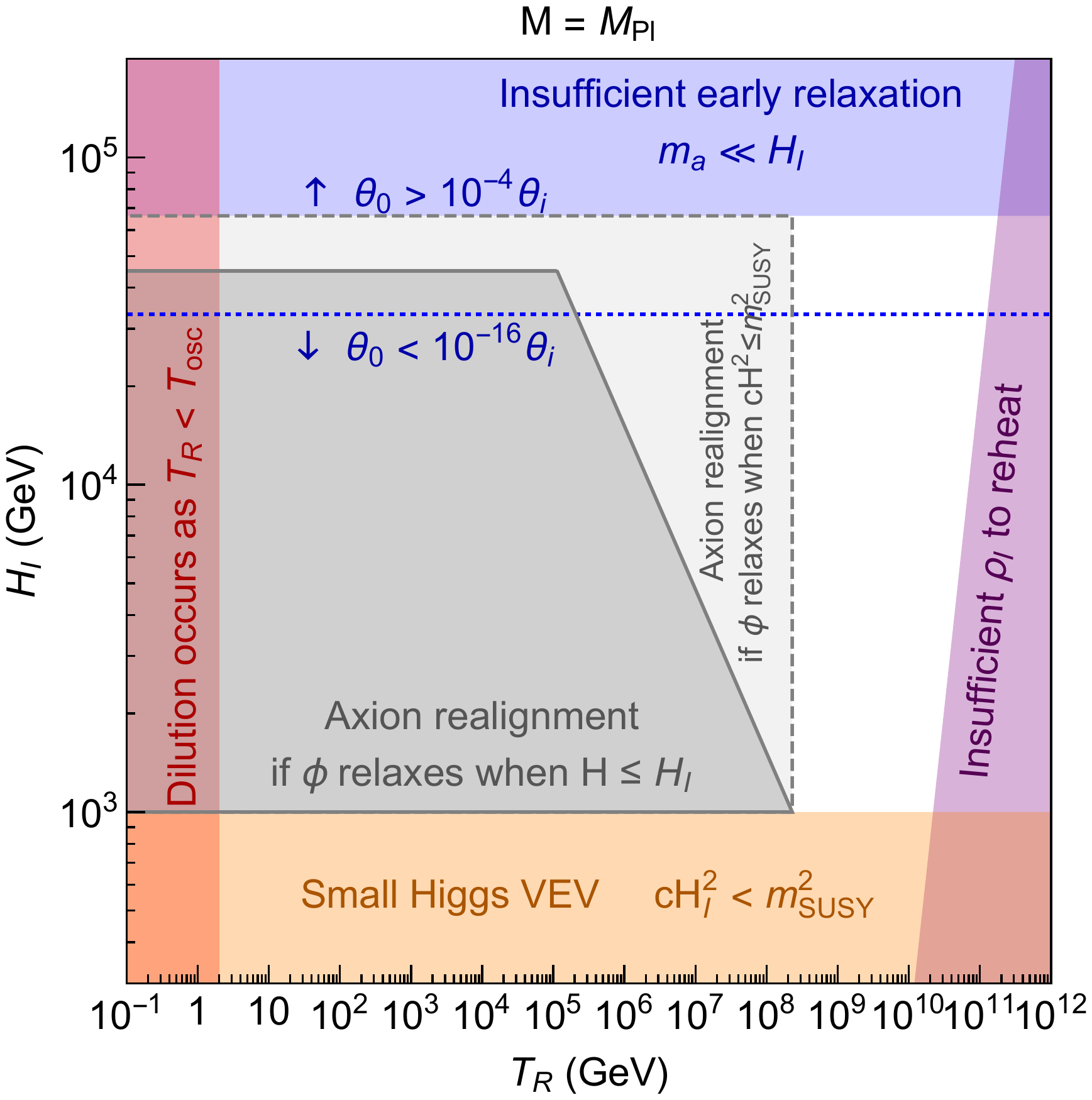}
	\caption{Parameter space for the inflationary Hubble scale $H_I$ and reheat temperature $T_R$ given $f_a = 3\times10^9 \, \GEV$, $m_{\tilde{g}}  = m_\text{SUSY} = \TEV$, $B\mu = m_\text{SUSY}^2/\tan\beta$, $\tan\beta = 50$, $N_e = 60$, and $\phi_i = M$. The left (right) panel is for the cutoff scale $M = M_{\rm GUT} \ (\mpl)$ respectively.}
\label{fig:HITR}
\end{figure}

The first consistency check we should perform is to ensure the axion mass is larger than Hubble friction during inflation. As extensively noted, for a given cutoff scale $M$, the large Higgs VEV $\phi_i \simeq M$ determines $\Lambda'_\text{QCD}$ and the axion mass during inflation is enhanced. The suppression of the angle by early relaxation can be approximated by Eqs.~(\ref{eq:theta_fric})~and~(\ref{eq:theta_relax}). As a benchmark point, we require the suppression factor to be $\theta_0/\theta_i =10^{-4}$ or smaller during the number of e-foldings of $60$. This gives an upper bound on the value of $H_I$, which is shown in the blue regions of Fig.~\ref{fig:HITR} with the left (right) panel for $M = M_{\rm GUT}\equiv 2\times 10^{16}~{\rm GeV} \ (\mpl)$ respectively. The blue contours are also shown for $\theta_0/\theta_i =10^{-16}$. The orange regions reflect a lower bound on the value of $H_I$ from requiring $cH_I^2 > \max(m_{\rm SUSY}^2, B\mu)$ necessary for obtaining a large Higgs VEV and the phase shift, respectively.

One needs to carefully consider the evolution of the axion potential after inflation ends. There must be a transition of the value of $\xi$ from the inflationary minimum toward today's minimum. This transition necessarily induces the transition of the minimum from $\pi$ to 0 in the axion potential.
If this transition occurs at a time when the enhanced axion mass is still comparable to or larger than Hubble, the misalignment angle could relax to a value very different from $\pi$. To understand this constraint, we turn to the post-inflationary evolution of the Higgs fields.

The Hubble induced mass terms which stabilize $\phi$ at a large VEV are tied to the inflaton energy density. If the sign of the Hubble induced terms remains the same after inflation, $\phi$ continues to be trapped around $M$.  This means the radial and angular directions of the Higgs fields do not oscillate until the Hubble induced terms become subdominant to the MSSM soft terms of the corresponding mode as the inflaton energy density redshifts and/or decays.

A second possibility for this evolution is that the sign of the Hubble induced mass flips after inflation (except for the $H_u H_d$ term.)
This may occur in two-field inflation models.
For example, with $K = ( - c_Z |Z|^2 +  c_{\bar{Z}} |\bar{Z}|^2 ) |\phi|^2$ with $c_Z > c_{\bar{Z}}$ and $W = m_Z Z \bar{Z}$,  we assume that the scalar component of $Z$ acquires a large field value and drives inflation. It is $\bar{Z}$ whose $F$-term, $F_{\bar{Z}}^2 = m_Z^2 \phi_{Z}^2$, is non-zero. During inflation, the kinetic energy of $\phi_Z$ is much smaller than its potential energy, i.e.~$|\partial_\mu Z|^2 \ll F_{\bar{Z}}^2$, and thus the Hubble induced mass for $\phi$ is negative. As inflation ends, $Z's$ potential and kinetic energies become comparable but, since $c_Z > c_{\bar{Z}}$, the sign of the Hubble induced mass for the Higgs radial mode flips to positive. Consequently, $\phi$ is no longer trapped at a large VEV but oscillates towards the origin immediately after inflation. The early onset of radial oscillations helps because a longer period of redshifting in $\phi$ suppresses $\Lambda_{\rm QCD}'$, which leads to the desired post-inflationary suppression of the axion mass.

It is only necessary to track the ratio $m_a/H$ between the onset of angular oscillations at $cH^2 = B\mu$ and thermalization of the Higgs fields. During this period, the Higgs phase $\xi$ can evolve and with it comes the shift in the axion potential. If the axion mass is subdominant to Hubble friction, however, the axion field is overdamped and remains agnostic to this evolution. When the Higgs is finally thermalized, its energy density is depleted and the field is quickly set to the minimum today, removing the axion mass enhancement. As a result, to preserve the prediction of the axion misalignment angle, $m_a/H$ needs to stay under unity during this period. 

Thermalization of the Higgs is mediated by scattering with gluons via a loop-suppressed operator \cite{Bodeker:2006ij, Mukaida:2012qn}
\begin{equation}
\label{eq:Gamma_h}
\Gamma_h = \frac{B}{16\pi^2}\frac{T^2}{\phi} ,
\end{equation}
with $B \simeq 10^{-2}$ and $\phi$ identified as the oscillation amplitude. Interestingly, due to the scaling properties during a matter-dominated era, the Higgs scattering generates a radiation energy density that is constant in time, whose contribution to thermal bath's temperature is
\begin{equation}
T_h = \left( \frac{30}{\pi^2 g_*(T_h)}   \frac{B}{16 \pi^2}  \right)^{1/2} \left( \frac{m_\phi^2 \phi_i}{H_I} \right)^{1/2}, 
\end{equation}
with $\phi_i$ as the field value of the Higgs at the end of inflation. This radiation persists throughout the evolution until the Higgs fields are thermalized at $H \simeq \Gamma_h$. This radiation is important because in some cases it can dominate over the radiation produced from the inflaton decay and cause a period of a constant temperature in the cosmological evolution. This has the effect of maintaining the finite temperature suppression to the axion mass 
\begin{equation}
m_a(T) = \frac{1}{4\pi}\frac{m_{\tilde{g}}^{1/2} \Lambda_\text{fid}^{3/2}}{f_a} \left(\frac{\Lambda'_\text{QCD}}{T}\right)^n ,
\end{equation}
where $n=3$ ($n=0$) for $T>\Lambda_\text{QCD}$ ($T<\Lambda_\text{QCD}$). The temperature dependence is determined by the contribution from the gauge multiplets, while the contribution from chiral multiplets vanishes because of the cancellation between the RGE contribution and the fermion mass suppression.
In the extended models, $\Lambda^{'}_\text{QCD}$ may be different from the estimate in Eq.~(\ref{eq:Lfid}) because of the backreaction from strong QCD dynamics. We note that the value of $m_a$ evolves after inflation not only due to this temperature suppression, but also its explicit dependence on $\Lambda_\text{fid}\propto \phi^{2/3}$, which evolves as the Higgs oscillation amplitude redshifts until the Higgs fields thermalize and settle into today's vacuum. “Including the decrease in $\Lambda^{'}_\text{QCD}$ from the redshift in $\phi \propto H^\frac{1}{1+w}$, with the equation of state of the total energy density $w$, the axion mass scales like $m_a|_\text{MD} \propto H^{\frac{3+2n}{3}}/T^n$ during the early matter-dominated era ($w=0$) and $m_a|_\text{RD} \propto H^{\frac{3+2n}{4}}/T^n$ during the radiation-dominated era ($w=1/3$) after reheating.” These considerations of the Higgs oscillations and axion mass suppression are taken into account when determining the post-inflationary constraint in the regions shaded in dark gray (light gray enclosed by the dashed contour) in Fig.~\ref{fig:HITR} assuming that the radial direction starts oscillation at $\sqrt{c}H \simeq m_{\rm SUSY}$ (right after inflation) respectively. This constraint is milder in the left panel because $M = M_{\rm GUT}$ starts out with a smaller axion mass during inflation than $M = \mpl$ so $m_a / H$ is more likely to be less than unity during the transition period. In fact, these gray regions disappear for $M \lesssim 10^{16}$ GeV opening up regions of low $T_R$ even though the blue constraint becomes stronger.

Another requirement of DAMP$_\pi$ comes from avoiding PQ symmetry restoration. In a thermal environment, the PQ breaking field saxion $P$ acquires a thermal mass $y^2T^2 P^2$ because of the Yukawa coupling $yPQ\bar{Q}$ with the PQ quarks $Q, \bar{Q}$. This thermal mass can be relevant at large temperatures and stabilize $P$ at a vanishing value to restore PQ symmetry. This can be easily prevented if the symmetry breaking is enforced by the following superpotential,
\begin{equation}
\label{eq:WP}
\Delta W_P = \lambda S \left(P\bar{P}-f_a^2 \right)  .
\end{equation}
The $F$-term of $S$ stabilizes $P$ and $\bar{P}$ in the moduli space $P\bar{P}=f_a^2$ to break PQ. The coupling constant $\lambda$  may be as large as $4\pi$ in strongly coupled models~\cite{Harigaya:2015soa,Harigaya:2017dgd}. To ensure that PQ is not thermally restored after inflation where the maximum temperature achieved during reheating is $T_{\rm max} \simeq (H_I \mpl T_R^2)^{1/4}$,%
\footnote{The actual maximal temperature is smaller after taking into account the efficiency of the thermalization~\cite{Harigaya:2013vwa}.}
the thermal mass must be less than $\lambda f_a < 4\pi f_a$ at this time, giving an upper bound on the Yukawa coupling
\begin{equation}
\label{eq:yPQresto}
y \lesssim 1.5 \left( \frac{m_{P,\bar{P}}}{f_a} \right) \left( \frac{f_a}{3\times 10^9 \ \GEV} \right) \left( \frac{\TEV}{H_I} \right)^{1/4} \left( \frac{10^{10} \, \GEV}{T_R} \right)^{1/2},
\end{equation}
which can be easily satisfied with $y \lesssim \mathcal{O}(1)$ in the allowed parameter space of Fig.~\ref{fig:HITR}.
The constraint is stronger if the mass of the PQ symmetry breaking field is only as large as $m_{\rm SUSY}$, but we do not pursue this issue further. 

A runaway potential for $P$ is generated by strong dynamics via $W_\text{eff} \simeq \left(1/4\pi\right)^2 \Lambda_\text{fid}^3 \left(P/f_a\right)^{1/3}$, which introduces an effective saxion mass of order $\left(1/4\pi\right)^2 \Lambda_\text{fid}^3 / f_a^2$. Keeping in mind the dependence of $\Lambda_\text{fid}$ on both the cutoff scale $M$ and $\tan\beta$ from Eq.~(\ref{eq:LQCD}), we find this effective saxion mass for the parameters in Fig.~\ref{fig:HITR} to be of order $10^2~\GeV$ ($10^6~\GeV$) for $M = M_\text{GUT}$  ($\mpl$) respectively. For values of $H_I$ in the allowed parameter space, a Hubble induced mass for $P$ is not always large enough to stabilize $P$. For simplicity, we stabilize $P$ by superpotential terms $W = m_P PY + m_{\bar{P}} \bar{P} \bar{Y}$. The $F$-terms of $Y$ and $\bar{Y}$ give large masses $m_P \sim m_{\bar{P}}$ to $P$ and $\bar{P}$. These masses can be as large as $4\pi f_a$ without destroying the moduli space $P\bar{P}= f_a^2$ and are large enough to stabilize $P$ against the runaway potential. The quarks $Q$ and $\bar{Q}$ have a large mass $yP$, which allows us to neglect the effects of their Hubble induced masses in the minimal models. We will see in Sec.~\ref{sec:cosmo_ext} that larger values of $\Lambda'_\text{QCD}$ and $H_I$ will modify the dynamics of these fields.

Finally, the purple regions in Fig.~\ref{fig:HITR} are excluded by energy conservation which restricts the reheat temperature $T_R$ to a maximum value dictated by the energy in the inflaton, $\rho_I \simeq \mpl^2 H_I^2 \gtrsim T_R^4$. In the red regions, the axion starts to oscillate from the hilltop towards today's minimum during a matter-dominated era by the inflaton, in which case reheating produces entropy, dilutes the axion abundance, and spoils the prediction of a small $f_a$. Even though the minimal model has proven to provide a viable cosmology for DAMP$_\pi$, we explore extended models in the following subsection to further broaden the parameter space.

\subsection{Extended Models}
\label{sec:cosmo_ext}

The most stringent constraint in the minimal model is a relatively low upper bound on $H_I$ due to difficulties in enhancing the axion mass during inflation. As shown in Eq.~(\ref{eq:LQCD}), we can enhance $\Lambda'_\text{QCD}$, and consequently $m_a$, by introducing additional matter content. As we raise the value of $\Lambda'_\text{QCD}$, the values of fields in the PQ sector may be shifted from the one in the vacuum and impact the evaluation of the axion mass. The field value of the Higgs may be also affected.

We consider a simple model where a PQ symmetry breaking field $P$ couples to KSVZ quarks $Q\bar{Q}$ by a Yukawa coupling $y$ \cite{Kim:1979if,Shifman:1979if}.
We need to reliably evaluate the VEVs of both $P$ and $Q\bar{Q}$, as both are PQ charged and the decay constant during inflation $f_I$ is given by the larger of $P$ and $(Q\bar{Q})^{1/2}$.
The superpotential of $P$ and $Q\bar{Q}$ is 
\begin{equation}
\label{eq:Wads}
W = \frac{1}{\left(4\pi\right)^3}\frac{\widetilde{\Lambda}^4}{\left(Q\bar{Q}\right)^{1/2}} + yPQ\bar{Q},
\end{equation}
where the first term is the non-perturbative Affleck-Dine-Seiberg superpotential~\cite{Affleck:1983mk}. The scale $\tilde{\Lambda}$ is related with $\Lambda_{\rm fid}$ via
\begin{equation}
\widetilde{\Lambda} = \Lambda_\text{fid} \left( \frac{\Lambda_\text{fid}}{y f_a} \right)^{1/8} .
\end{equation}
The relation can be obtained by comparing the effective superpotential after integrating out $Q\bar{Q}$ from Eq.~(\ref{eq:Wads}) and the effective potential
\begin{equation}
\label{eq:Weff}
W_\text{eff} \simeq \frac{1}{\left(4\pi\right)^2}\Lambda_\text{fid}^3 \left(\frac{P}{f_a}\right)^{1/3} ,
\end{equation}
where the effect of a $P$ field value different from $f_a$ is included.
Note that the potential of $P$ from strong dynamics exhibits a runway behavior, $\partial W_\text{eff} /\partial P \propto P^{-2/3}$. For this reason, we introduce a higher dimensional term $|P|^6/M^2$ to further stabilize $P$ at a large field value which can come from a superpotential term $(\chi/M)P^3$, where $\chi$ is a chiral field.
We also consider the Hubble induced mass of $P$ and $Q\bar{Q}$. Explicitly we take
\begin{equation}
\label{eq:VforPQ}
\Delta V \simeq cH_I^2|Q|^2 + cH_I^2|\bar{Q}|^2 - cH_I^2|P|^2 +  \frac{|P|^6}{M^2}.
\end{equation}
We stress that these terms are used in our analysis but they are not the only possible extensions to DAMP$_\pi$.
The sign of the Hubble induced mass of $Q$ and $\bar{Q}$ is taken to be positive to ensure that $Q$ is not destabilized by the Hubble induced mass. We study a negative Hubble induced mass for $P$ for the following reason. We find that $Q>P$ and hence $f_I$ is dominated by $Q$ for this choice of the signs. If the Hubble induced mass of $P$ is positive instead, the field value of $P$ becomes smaller and makes the field value of $Q$ larger because of the smaller mass of $Q$. This increases $f_I$ while decreasing the dynamical scale and suppressing the axion mass.

As the Hubble induced mass of $Q$ breaks the supersymmetry of the QCD sector, the axion mass may non-trivially depends on the parameters.
When $yP$ is larger than $Q$, $Q\bar{Q}$ can be integrated out at the mass threshold $yP$ and the effective potential is given by Eq.~(\ref{eq:Weff}). The axion mass is given by
\begin{align}
\label{eq:ma_yP}
m_a \simeq \frac{1}{4\pi}\frac{m_{\tilde{g}}^{1/2} \Lambda_\text{fid}^{3/2}}{f_I} \left( \frac{P}{f_a} \right)^{1/6}.
\end{align}
If $Q$ is larger than $yP$, the theory below the mass threshold $Q$ is a supersymmetric, pure $SU(2)$ gauge theory with a dynamical scale $\tilde{\Lambda} (\tilde{\Lambda}/Q)^{1/3}$. The axion mass is then given by
\begin{align}
\label{eq:ma_Q}
m_a \simeq  \frac{1}{4\pi} \frac{m_{\tilde{g}}^{1/2} \widetilde{\Lambda}^{3/2}}{f_I} \left( \frac{\widetilde{\Lambda}}{4\pi Q} \right)^{1/2} =  \frac{1}{4\pi} \frac{m_{\tilde{g}}^{1/2} \Lambda_\text{fid}^{3/2}}{f_I}  \left(\frac{\Lambda_{\rm fid}^3}{16 \pi^2 y f_a Q^2}\right)^{1/4}.
\end{align}
Note that the two formulae agree with each other when the field value of $Q$ is determined by the $F$-term condition of $Q\bar{Q}$ from the superpotential in Eq.~(\ref{eq:Wads}).

Strong dynamics also affects the Higgs. The effective superpotential of $\phi$ is given by
\begin{align}
W\simeq \frac{\Lambda_\text{eff}^3}{16\pi^2} \left( \frac{\phi}{M} \right)^2,~~
\Lambda_\text{eff}^3 =
\begin{cases}
\Lambda_\text{fid}^3 \left(\frac{P}{f_a}\right)^{1/3} & yP >Q \\
\Lambda_\text{fid}^3  \left(\frac{\Lambda_{\rm fid}^3}{16 \pi^2 y f_a Q^2}\right)^{1/2} & yP < Q.
\end{cases}
\end{align}
This  gives the Higgs a mass $\simeq \Lambda_\text{eff}^3/(16\pi^2 M^2)$, which should be smaller than $\sqrt{c} H_I$.

\begin{figure}
	\includegraphics[width=0.495\linewidth]{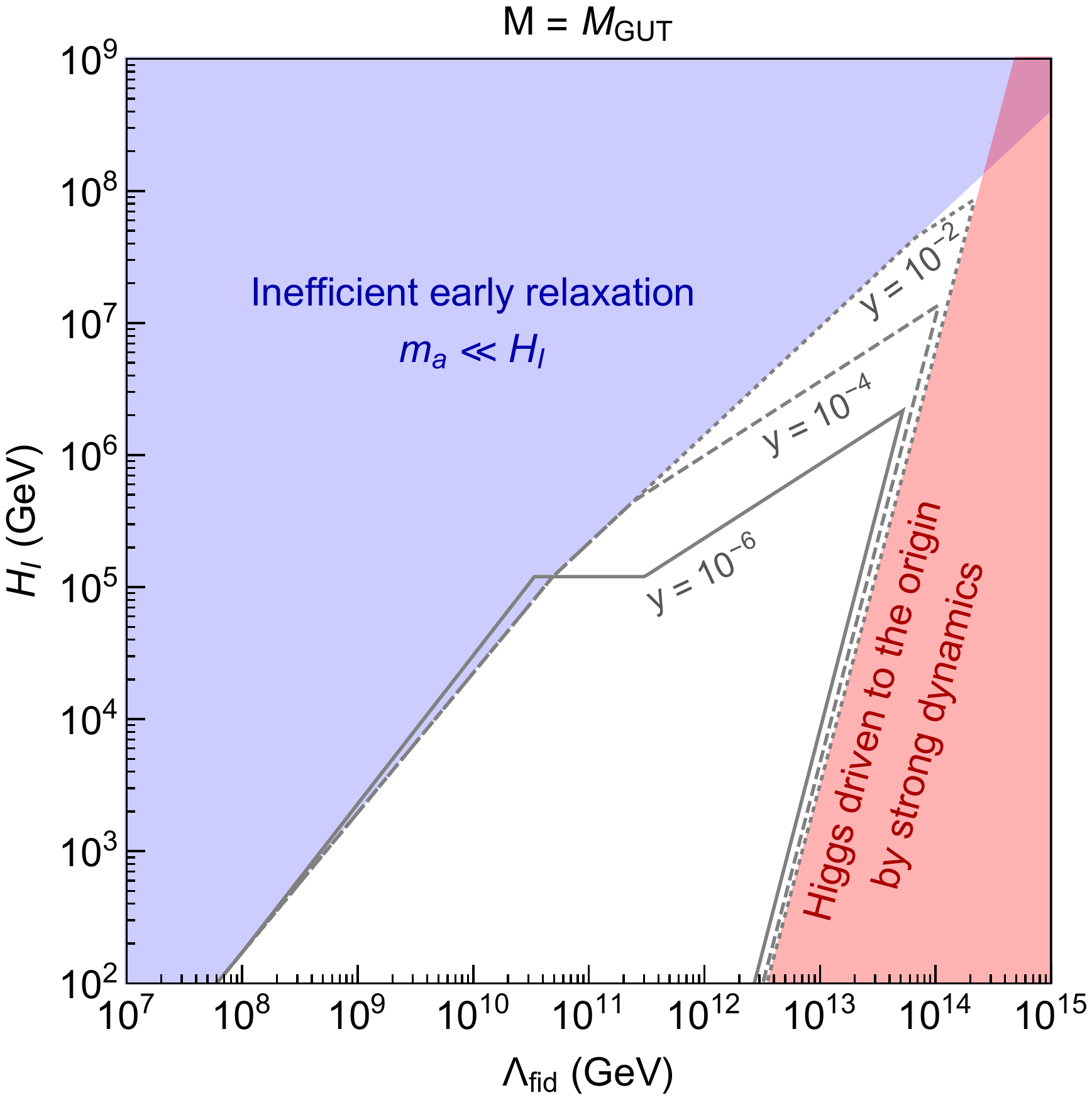}	\includegraphics[width=0.495\linewidth]{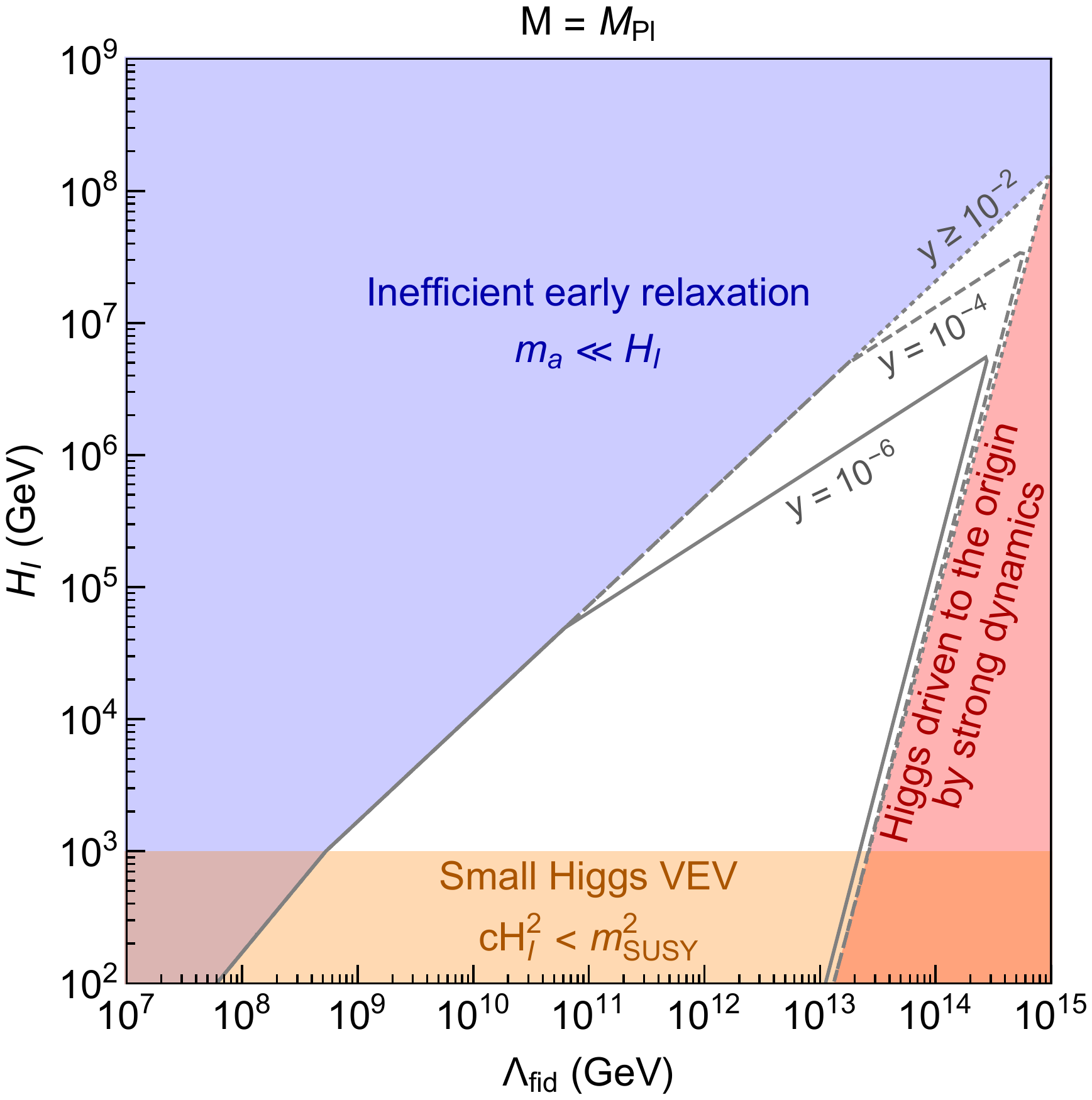}
	\caption{Parameter space for the inflationary Hubble scale $H_I$ and the fiducial confinement scale $\Lambda_{\rm fid}$ defined in Eq.~(\ref{eq:Lfid}) given $f_a = 3\times10^9 \, \GEV$, $m_{\tilde{g}} = m_\text{SUSY} = \TEV$, and $\phi_i = M$. The left (right) panel is for the cutoff scale $M = M_{\rm GUT} \ (\mpl)$ respectively.}
	\label{fig:HIvsLambdaI}
\end{figure}

By computing and comparing the axion mass to $H_I$, we put an upper bound on the allowed values of $H_I$ such that DAMP$_\pi$'s first criterion is fulfilled during inflation, which is shown in Fig.~\ref{fig:HIvsLambdaI}. In deriving the blue-shaded region, we integrate out $Q \bar{Q}$, obtain the scalar potential of $P$ from the effective potential Eq.~(\ref{eq:Weff}), add the potential of $P$ in Eq.~(\ref{eq:VforPQ}), determine the field value of $P$ during inflation, and compute the axion mass. This corresponds to the case where $Q$ is actually determined by the $F$-term condition $\partial W / \partial Q = 0$. The gray contours show the constraint using the full potential described above. As $y$ becomes larger the constraints approach to the blue-shaded region. An additional constraint shown in the red regions arises because strong dynamics drives the Higgs to the origin.

In computing the field values of $P$ and $Q$, we treat $Q$ as a canonically normalized field. This is a good approximation if $yP$ or $Q$ is above the dynamical scale. We find that in the allowed parameter space, either $yP$ or $Q$ is no smaller than one order of magnitude below the dynamical scale so we expect the approximation gives a good order of magnitude estimate.

We now discuss the post-inflationary evolution and constraints similar to Sec.~\ref{subsec:cosmo_min}. Although this study is comprehensive in evaluating the inflationary constraints, the thermal masses for $P$ and $Q$ dramatically complicate PQ dynamics during and after reheating. Nonetheless, we are able to identify a large allowed parameter space in the $H_I$, $T_R$ plane in the following way. There exists a wide region where the Higgs fields thermalize before the Higgs angular mode $\xi$ begins to oscillate so that $\Lambda'_\text{QCD}$ is quickly set to today's value $\Lambda_\text{QCD}$ before $\xi$ has a chance to evolve and shift the axion potential. Physically, this means that the axion potential turns off before the location of its minimum shifts. Therefore, the prediction of the misalignment angle is automatically preserved without the need to track the post-inflationary evolution of $P$ and $Q\bar{Q}$. This is the case when the Higgs scattering rate in Eq.~(\ref{eq:Gamma_h}) equals the Hubble rate before $cH^2$ drops below $B\mu$. This region is described by an allowed window of $T_R$ for a given $H_I$
\begin{align}
\label{eq:pretherm}
T_R & \gtrsim 10^8 \ \GeV \ \left( \frac{M}{M_{\rm GUT}} \right)^{3/2} \left( \frac{B\mu}{\TEV} \right)^{3/4}  \left( \frac{\phi_i}{M_{\rm GUT}} \right) \left( \frac{10^5 \ \GEV}{H_I} \right) , \\
T_R & \lesssim 6\times10^{11} \ \GeV \ \left( \frac{M_{\rm GUT}}{M} \right)^{3/2} \left( \frac{\TEV}{B\mu} \right)^{3/4} \left( \frac{H_I}{10^5 \ \GEV} \right)^2 \left( \frac{M_{\rm GUT}}{\phi_i} \right)^2  ,
\end{align}
where the two distinct formulae come from thermalization during the matter- and radiation-dominated epochs respectively. This window becomes wider as $H_I$ increases so as long as 
\begin{equation}
H_I > 6 \ \TEV \left( \frac{M}{M_{\rm GUT}} \right) \left( \frac{B\mu}{\TEV} \right)^{1/2} \left( \frac{\phi_i}{M_{\rm GUT}} \right) 
\end{equation}
a consistent range of $T_R$ exists. The upper bound on $H_I$ ultimately enters from the inflationary constraint shown in the blue regions of Fig.~\ref{fig:HIvsLambdaI}.

We now comment on one plausible extension to further open up the parameter space with higher $H_I$. The gluino mass affects the axion mass as in Eq.~(\ref{eq:ma}) and is assumed to stay invariant between inflation and today. If $m_{\tilde{g}}$ is also larger during inflation, the axion mass and the upper bound on $H_I$ can be raised by as much as $(\Lambda_{\rm fid} / m_{\tilde{g}})^{1/2}$, making high scale inflation easily compatible with DAMP$_\pi$.

With respect to PQ restoration in the extended models, we can assume the same PQ breaking mechanism as in Sec.~\ref{subsec:cosmo_min} so the constraint in Eq.~(\ref{eq:yPQresto}) applies equally here.

While the post-inflationary constraints are not fully evaluated, these extended models have been shown capable of fulfilling the criteria of DAMP$_\pi$ while extending the allowed parameter space to much higher $H_I$ than in Sec.~\ref{subsec:cosmo_min}.

\section{Conclusion}
\label{sec:conclusion}
It has been widely known that the misalignment mechanism can source axion dark matter in the early universe with a decay constant $f_a \simeq \mathcal{O}(10^{12})$ GeV. For $f_a \ll 10^{12}$~GeV as is of interest to many experimental searches, the observed DM abundance can be obtained if the misalignment angle $\theta_{\rm mis}$ is taken sufficiently close to $\pi$, where the anharmonic effect becomes important. In particular, when the axion is very close to the hilltop of the potential, the onset of oscillations is delayed so the axion abundance is less redshifted and thus more enhanced. As demonstrated in Fig.~\ref{fig:fa} and Eq.~(\ref{eq:fa_F_anh}), $f_a \simeq 10^{10}$ GeV corresponds to $\delta\theta \equiv \pi - \theta_{\rm mis} \simeq \mathcal{O}(10^{-3})$, while $f_a \simeq 4\times10^9$ GeV already requires $\delta \theta \simeq \mathcal{O}(10^{-9})$. Such a small $\delta\theta$ has generically been understood as fine-tuning of the initial condition. In this paper, we offer an explanation to this small $\delta\theta$ using axion dynamics in the early universe. 

We point out that a class of models violates the canonical assumption that the axion field is overdamped by Hubble friction and takes a random value during inflation. Instead, there exist numerous possibilities wherein the axion is large compared to Hubble during inflation and thus relaxes to the minimum of the potential. We refer to this mechanism as Dynamical Axion Misalignment Production (DAMP). Additionally, if the model possesses an approximate CP symmetry, then the axion potential may receive a phase shift of $\pi$ because the nearly real parameters for setting the axion minimum can flip the sign between inflation and the QCD phase transition. This shift converts the potential minimum into a maximum and explains why the axion is very close to the hilltop---a mechanism we dub DAMP$_\pi$. 

We explicitly construct models for DAMP$_\pi$, where the higher dimensional coupling between the Higgs in the MSSM and the inflaton gives rise to a large axion mass and the phase shift of the axion potential. Specifically, a negative Hubble induced mass drives the Higgs to a large field value that enhances the quark masses, which in turn raise the QCD scale. The axion is larger than usual due to stronger QCD dynamics. Lastly, a Hubble induced $B\mu$ term that carries an opposite sign from that of the MSSM necessarily induces a shift in the axion potential by $\pi$. Together, renormalization from the SM Yukawa couplings and any additional CP violating phases in the model can provide the desired finite phase shift between $\mathcal{O}(10^{-16}\textrm{--}10^{-3})$. This example works only if the number of generations is odd.

Strikingly, due to the anharmonic effects of the axion potential, the prediction of $f_a$ from the DM abundance has an extraordinarily mild logarithmic dependence on $\delta \theta \ll 1$. Therefore, DAMP$_\pi$ makes a rather sharp prediction of $3 \times 10^9~\GEV \lesssim f_a \lesssim 10^{10}$~GeV. Other phenomenological features of DAMP$_\pi$ are as follows. Due to early relaxation, the fluctuation of the axion field is exponentially damped and hence dark matter isocurvature perturbations are suppressed.
% dark matter has exponentially damped isocurvature perturbations and instead has nearly pure adiabatic perturbations.
The upper bound on $H_I$ from isocurvature perturbations does not apply.
%on $H_I$ do not apply to this model, so $H_I$ has no upper bound.

The impact of this anharmonicity on the structure formation has been investigated in the literature. Refs.~\cite{Turner:1985si,Strobl:1994wk} study this numerically and show that isocurvature perturbation modes whose wavelengths are larger than the horizon size are enhanced by anharmonic effects. Refs.~\cite{Kolb:1993hw,Kolb:1993zz} numerically study the anharmonicity effects on the growth of structures arising from the large fluctuations in an inhomogeneous background, i.e.~in the context of post-inflationary PQ symmetry breaking. Finally, Ref.~\cite{Greene:1998pb} has shown that in a quasi-homogeneous region, parametric resonance can be important for amplifying fluctuations, and this effect is monotonically enhanced for larger misalignment angles. While the anharmonicity effect may stimulate structure formation, the aforementioned works are not directly applicable to this model as the isocurvature perturbations are suppressed and the PQ symmetry is already broken during inflation.

%We note that this anharmonicity may also impact structure formation. Refs.~\cite{Turner:1985si,Strobl:1994wk} study this numerically and show that isocurvature perturbation modes whose wavelengths are larger than the horizon size are enhanced by anharmonic effects. Refs.~\cite{Kolb:1993hw,Kolb:1993zz} numerically study the anharmonicity effects on the growth of structures arising from the large fluctuations in an inhomogeneous background, i.e.~in the context of post-inflationary PQ symmetry breaking. While the anharmonicity effect may stimulate structure formation, the aforementioned works are not directly applicable to this model as the fluctuations from isocurvature perturbations and post-inflationary PQ breaking are both absent. Finally, Ref.~\cite{Greene:1998pb}  has shown that in a quasi-homogeneous region, parametric resonance can be important for amplifying fluctuations, and this effect is monotonically enhanced for larger misalignment angles. This implies that with a suitable seed of perturbations, structure formation might be enhanced in our scenario.

The axions from the misalignment mechanism are necessarily cold---a feature to distinguish from other non-thermal production mechanisms. It is also potentially interesting to study the imprints of maximal CP violation on the QCD phase transition as well as Big Bang nucleosynthesis.

 {\bf Acknowledgment.}---%
The authors thank Asimina Arvanitaki, Joshua W. Foster, Andrew J. Long, Aaron Pierce, Benjamin R. Safdi, and Ken van Tilburg for discussions. K.H. thanks the Leinweber Center for Theoretical Physics for the warm hospitality he received during his visit when most of this work was done. The work was supported in part by the DoE Early Career Grant DE-SC0019225 (R.C.) and the DoE grant DE-SC0009988 (K.H.).

\bibliography{bibtexrefs}

\end{document}